\documentclass{pasj00}
\begin{document}
\SetRunningHead{OGURI ET AL.}{SDSS J1335+0118}
\Received{}
\Accepted{}
\title{SDSS J1335+0118: A New Two-Image Gravitational Lens}
\author{%
Masamune \textsc{Oguri},\altaffilmark{1}
Naohisa \textsc{Inada},\altaffilmark{1}
Francisco J. \textsc{Castander},\altaffilmark{2}
Michael D. \textsc{Gregg},\altaffilmark{3,4}\\
Robert H. \textsc{Becker},\altaffilmark{3,4}
Shin-Ichi \textsc{Ichikawa},\altaffilmark{5}
Bartosz \textsc{Pindor},\altaffilmark{6}
Jonathan \textsc{Brinkmann},\altaffilmark{7}\\
Daniel J. \textsc{Eisenstein},\altaffilmark{8}
Joshua A. \textsc{Frieman},\altaffilmark{9,10}
Patrick B. \textsc{Hall},\altaffilmark{6}
David E. \textsc{Johnston},\altaffilmark{9}\\
Gordon T. \textsc{Richards},\altaffilmark{6}
Paul L. \textsc{Schechter},\altaffilmark{11}
Donald P. \textsc{Schneider},\altaffilmark{12}
and
Alexander S. \textsc{Szalay}\altaffilmark{13}
}
\altaffiltext{1}{Department of Physics, University of Tokyo, Hongo
7-3-1, Bunkyo-ku, Tokyo 113-0033.} 
\altaffiltext{2}{Institut d'Estudis Espacials de Catalunya/CSIC,
Gran Capita 2-4, 08034 Barcelona, Spain.}
\altaffiltext{3}{Department of Physics, University of California at
Davis, \\ 1 Shields Avenue, Davis, CA 95616, USA.}
\altaffiltext{4}{Institute of Geophysics and Planetary Physics, 
Lawrence Livermore National Laboratory, \\ L-413, 7000 East Aveneu,
Livermore, CA 94550, USA.} 
\altaffiltext{5}{National Astronomical Observatory, 2-21-1 Osawa,
Mitaka, Tokyo 181-8588.} 
\altaffiltext{6}{Princeton University Observatory, Peyton Hall,
Princeton, NJ 08544, USA.} 
\altaffiltext{7}{Apache Point Observatory, P.O. Box 59, Sunspot,
NM88349, USA.}
\altaffiltext{8}{Steward Observatory, University of Arizona,\\ 933 North
Cherry Avenue, Tucson, AZ 85721, USA.} 
\altaffiltext{9}{Astronomy and Astrophysics Department, University of
Chicago, \\ 5640 South Ellis Avenue, Chicago, IL 60637, USA.}
\altaffiltext{10}{Fermi National Accelerator Laboratory, P.O. Box
500, Batavia, IL 60510, USA.}
\altaffiltext{11}{Department of Physics, Massachusetts Institute of
Technology, \\ 77 Massachusetts Avenue, Cambridge, MA 02139, USA.} 
\altaffiltext{12}{Department of Astronomy and Astrophysics, Pennsylvania
State University, \\ 525 Davey Laboratory, University Park, PA 16802.}
\altaffiltext{13}{Department of Physics and Astronomy, Johns Hopkins
University, \\ 3701, San Martin Drive, Baltimore, MD 21218, USA.}

\maketitle

\begin{abstract}
We report the discovery of the two-image gravitationally lensed quasar
SDSS~J1335+0118. The object was selected as a lens candidate from the
Sloan Digital Sky Survey. The imaging and spectroscopic follow-up
observations confirm that the system exhibits two gravitationally lensed
images of a quasar at $z=1.57$. The image separation is $1\farcs56$.  We
also detect an extended component between the two quasar images, likely
the lensing galaxy. Preliminary mass modeling predicts the differential
time delay $\Delta t\sim 30h^{-1}{\rm day}$ assuming the redshift of the
lens galaxy is $0.5$. 
\end{abstract}

\section{Introduction}

Multiple images of a quasar produced by a foreground galaxy are useful 
for cosmological and astrophysical applications, including the
measurement of the cosmological constant through the lensing rate
\citep{turner90,fukugita90,chae02}, the determination of global Hubble
constant from time delays between images (Refsdal 1964, 1966),
the study of the mass distributions 
of the lensing galaxies 
\citep{kochanek91,metcalf01,chiba02,dalal02,schechter02,koopmans03,rusin03}, 
and the formation and evolution of galaxies
\citep{kochanek00,kochanek01,keeton01a,oguri02,ofek03}. 
A large homogeneous sample of lensed quasars with a well-understood 
selection function is essential in these statistical studies.
In addition, lensed quasars can be a powerful tool to study quasar
itself through microlensing, spectra, or variabilities of lensed quasars
\citep{chang84,grieger91,mediavilla98,mineshige99,yonehara99,wyithe00,yonehara01,lewis03}.
Therefore, it is important to make a large sample of lensed quasars also to
find more gravitational lens systems which are suitable for studying quasar.

The Sloan Digital Sky Survey (SDSS; \cite{york00,stoughton02,abazajian03})
has the potential to provide such a lensed quasar sample. The SDSS is a
survey to image $10^4{\rm deg^2}$ of the sky as well as to obtain
spectra of galaxies and quasars from the imaging data. The dedicated 2.5
meter telescope at Apache Point Observatory is equipped with a multi-CCD
camera \citep{gunn98} with five broad band centered at $3561$, $4676$,
$6176$, $7494$, and $8873${\,\AA} \citep{fukugita96b}. The imaging data
are automatically reduced by the photometric pipeline \citep{lupton01}.
The astrometric positions are accurate to about $0\farcs1$ for sources
brighter than $r=20.5$ \citep{pier03}. The photometric errors are
typically less than 0.03 magnitude \citep{hogg01,smith02}. The SDSS
quasar selection algorithm is presented in \citet{richards02}; the SDSS
 spectra cover $3800$--$9200${\,\AA} at a resolution of $1800$--$2100$.
The final spectroscopic quasar sample is expected to comprise $10^5$
quasars, thus will contain $\sim10^2$ lensed quasars given the typical
lensing probability $0.1\%$ \citep{turner84}.  Indeed, several new
gravitational lens systems have been discovered using SDSS data
(\authorcite{inada03a} 2003a,b,2004; \cite{morgan03}; \cite{johnston03};
\cite{pindor04}; \cite{oguri04}).  

We report the discovery of the lensed quasar SDSS~J133534.79+011805.5
(SDSS~J1335+0118) in the SDSS. This quasar has already been identified
in the Large Bright Quasar Survey (LBQS 1333+0133; \cite{hewitt91}), but
not as lensed. A search for lens candidates in the LBQS was carried out
by \citet{hewett98}, but those authors state that their search is not
sensitive to lens systems with $\leq 3''$ separations. The quasar was
selected as a lens candidate in the course of an ongoing search for
strongly lensed quasars in the SDSS. From the results of photometric
and spectroscopic follow-up observations, we conclude that the quasar is
lensed by an intervening galaxy. 

\section{Observations\label{sec:obs}}

\subsection{Candidate Selection}

\begin{figure}[t]
   \begin{center}
      \FigureFile(80mm,50mm){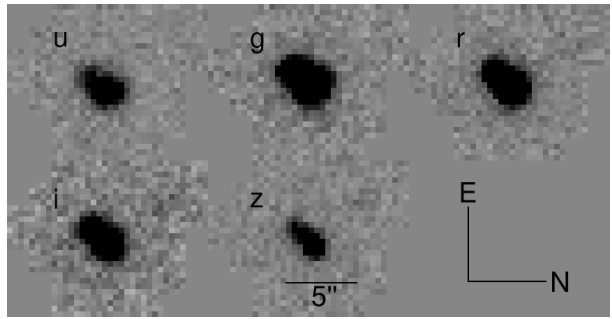}
   \end{center}
   \caption{SDSS images of SDSS~J1335+0118 in all bands. The image scale is
$0\farcs396{\rm pixel^{-1}}$. \label{fig:atlas}}
\end{figure}

To select SDSS lensed quasar candidates, we examine all
spectroscopically confirmed quasars with $z>0.6$. Quasars at smaller
redshifts have much lower probability of being lensed and also are often
extended because of their host galaxies, making it more difficult to
select lens candidates with our candidate selection algorithm described
below. We search for lens candidates using a combination of SDSS
parameters: {\tt dev\_L} (the likelihood parameter of fitting by de
Vaucouleurs profile), {\tt exp\_L} (the likelihood parameter of fitting
by exponential disk), and {\tt star\_L} (the likelihood parameter of
fitting by point spread function). This candidate selection algorithm
has already found three new lensed quasar systems (\authorcite{inada03a}
2003a,2004; \cite{pindor04}), and can identify gravitationally lensed
quasars with image separations $1\farcs0-2\farcs5$ quite well (N.~Inada
et al., in preparation). The SDSS $ugriz$ imaging data of
SDSS~J1335+0118 (see Figure \ref{fig:atlas}) clearly shows that the
system consists of two stellar components with similar colors, making it
an excellent lensing candidate. The total magnitudes of this system in
the SDSS photometric data are $u=18.21\pm0.02$, $g=17.83\pm0.01$,
$r=17.62\pm0.01$, $i=17.26\pm0.01$, and $z=17.14\pm0.03$. 

\subsection{Additional Imaging Observations}

\begin{figure}[t]
   \begin{center}
      \FigureFile(80mm,50mm){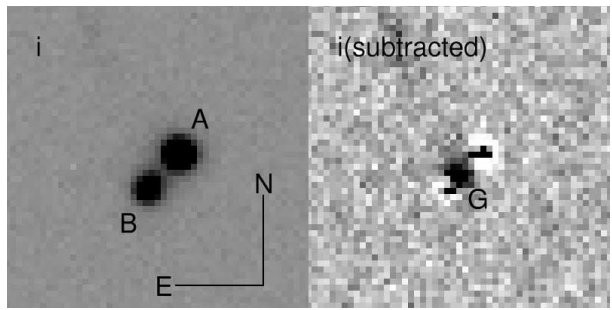}
   \end{center}
   \caption{Subaru Suprime-cam images of SDSS~J1335+0118 in the $i$-band.
 The image scale is $0\farcs2{\rm pixel^{-1}}$. {\it Left}: original
 Suprime-cam image. Components A and B are lensed quasar components.
 {\it Right}: PSF-subtracted image. The component G is likely to be the
 lens galaxy. \label{fig:1335_subaru}}
\end{figure}

We obtained a deep $i$-band image on 2003 May 28 with the Subaru Prime
Focus Camera (Suprime-Cam; \cite{miyazaki02}) of the National
Astronomical Observatory of Japan's Subaru 8.2-m telescope, which is
shown in Figure \ref{fig:1335_subaru}. The exposure time was 30 seconds
and the seeing was $0\farcs5-0\farcs6$. The image scale is $0\farcs2{\rm
pixel^{-1}}$. Each frame was bias-subtracted and flat-field corrected. 
Subtraction of the two quasar images using a nearby star as a
point-spread function (PSF) template reveals an extended object, denoted
as G, located on the line between A and B and closer to B (see Figure
\ref{fig:1335_subaru}). This configuration is expected for a standard
simple lens model such as singular isothermal sphere model, so component
G is likely to be the lens galaxy. In Table \ref{table:posflux}, we
summarize the position of each component and the relative fluxes of the
quasar components that are measured using Subaru Suprime-cam data.
The errors of the galaxy position is determined from Gaussian fits.

\begin{table*}
  \caption{Positions and flux ratios of SDSS~J1335+0118 in the
 Subaru $i$-band image\label{table:posflux}}
  \begin{center}
    \begin{tabular}{crrc}
     \hline\hline
      Object & $x$[arcsec]\footnotemark[$*$] &
     $y$[arcsec]\footnotemark[$*$] & Flux[arbitrary]\footnotemark[$\dagger$]\\
     \hline
     A & $0.000\pm0.001$   & $0.000\pm0.001$  & $1.0\pm0.2$ \\
     B & $-1.038\pm0.002$  & $-1.165\pm0.002$ & $0.374\pm0.075$ \\
     G & $-0.769\pm0.011$  & $-0.757\pm0.011$ &  $\cdots$\\
    \hline
     \multicolumn{4}{@{}l@{}}{\hbox to 0pt{\parbox{180mm}{\footnotesize
       \footnotemark[$*$] The positive directions of $x$ and $y$ are defined by
 West and North, respectively.
       \par\noindent
       \footnotemark[$\dagger$] Errors are broadened to 20\% to account
     for possible systematic effects. 
     }\hss}}
\end{tabular}
  \end{center}
\end{table*}

\begin{figure}[t]
   \begin{center}
      \FigureFile(80mm,50mm){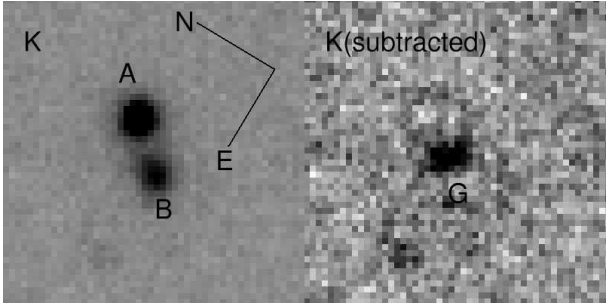}
   \end{center}
   \caption{Same as Figure \ref{fig:1335_subaru}, but the $K$-band images
 taken with NIRC on Keck I telescope are shown.  The image scale is
 $0\farcs15{\rm pixel^{-1}}$. \label{fig:1335_keck}}
\end{figure}

We also obtained $J$ and $K$-band data on 2003 April 19 with the Near
InfraRed Camera (NIRC; \cite{matthews94}) of the Keck~I telescope at
the W. M.  Keck Observatory on Mauna Kea, Hawaii, USA.  The total
exposure time was 900 seconds, seeing was variable, averaging $\sim
0\farcs6$.  Conditions were not photometric, with about 1 magnitude of
variable cloud.  The image scale is $0\farcs15{\rm pixel^{-1}}$.
Standard reduction procedures were followed to remove the dark current
and to flat field the data.  We show the $K$ image in Figure
\ref{fig:1335_keck}.  Although the galaxy component can be seen in the
direct image, we subtracted the PSF of two quasar components by
fitting a Gaussian PSF at each quasar location, revealing the lensing
galaxy more clearly (see Figure \ref{fig:1335_keck}).  As in the
$i$-band image shown in Figure \ref{fig:1335_subaru}, the galaxy
component is seen in the PSF-subtracted $K$-band image.  We compare the
distance between components A, B, and G in Table \ref{table:keckpos}.
The differences of the distances are large, but only $\sim 0\farcs1$ at
most and still much smaller than seeing sizes and pixel sizes.  The
position of the lens galaxy is particularly uncertain in the 
$K$ data because not only is the lensing galaxy relatively faint and
extended, but also because the seeing was quite variable on the night
the IR data were obtained and no suitable PSF star is present in the
NIRC field, making it difficult to properly subtract the PSF wings which
overlap the galaxy. Thus it may be possible that such residuals
systematically affect the estimated positions. We note that the system
is detected (but not resolved) by the Two-Micron All-Sky Survey (2MASS)
with $J = 16.23\pm0.10$, $H=15.47\pm0.11$, and $K=15.29\pm0.18$.  

\begin{table*}
  \caption{Comparison of positions between Subaru $i$-band and Keck
 $K$-band images\label{table:keckpos}}
  \begin{center}
    \begin{tabular}{ccc}
     \hline\hline
      Objects & $i$-band Distance[arcsec] &
     $K$-band Distance[arcsec] \\
     \hline
     AB & $1.560\pm0.002$ & $1.503\pm0.002$  \\
     AG & $1.079\pm0.011$ & $0.969\pm0.012$  \\
     BG & $0.489\pm0.011$ & $0.542\pm0.012$  \\
    \hline
    \end{tabular}
  \end{center}
\end{table*}

\subsection{Spectroscopic Observations}

\begin{figure}[t]
   \begin{center}
      \FigureFile(80mm,50mm){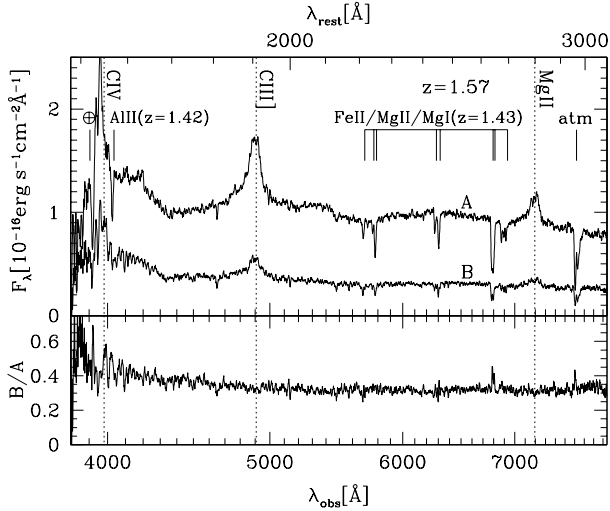}
   \end{center}
   \caption{ESO NTT spectra of SDSS~J1335+0118 components A and B. 
 The redshifts of components A and B estimated from Mg \emissiontype{II}
 emission lines are $z=1.57\pm0.03$ and $z=1.57\pm0.05$, respectively.
 Both components have associated C \emissiontype{IV} absorption at a
  blueshift of $2500{\rm km\,s^{-1}}$ from the quasar (marked by
 $\oplus$). In addition, both components have strong Fe
 \emissiontype{II}/Mg \emissiontype{II}/Mg \emissiontype{I} absorption
 system at $z=1.43$, and Al \emissiontype{II} absorption system at
 $z=1.42$. Note that the absorption at $\sim 7600${\AA} is atmospheric. 
 The bottom panel shows the ratio of spectra. \label{fig:spec}}
\end{figure}

The spectroscopic observation was done on 2003 March 2 with the red CCD
of the ESO Multi-Mode Instrument (EMMI) on the ESO New Technology
Telescope (NTT).  We used grism \#3 ($360\,{\rm line\,mm^{-1}}$,
$2.3${\,\AA}${\rm pixel^{-1}}$, covering $3850-8450${\,\AA}).  The
seeing was $1\farcs0$. The exposure time was 900 sec.  The slit was
positioned to obtain spectra of the two quasar components (A and B)
simultaneously.  The two components were clearly separated in the EMMI
image, making extraction simple. The data were reduced in a standard way
using IRAF\footnote{IRAF is distributed by the National Optical
Astronomy Observatories, which are operated by the Association of
Universities for Research in Astronomy, Inc., under cooperative
agreement with the National Science Foundation.}. The spectra are shown
in Figure \ref{fig:spec}. We find that both components A and B have C
\emissiontype{IV}, C \emissiontype{III]}, and Mg \emissiontype{II}
emission lines at the same wavelengths. The redshifts of components A
and B estimated from Mg \emissiontype{II} emission lines are
$z=1.57\pm0.03$ and $z=1.57\pm0.05$, respectively. We cross-correlate
the two spectra and estimate the velocity difference between two quasar
components to be $20\pm800{\rm km\,s^{-1}}$. We also plot the ratio of
the fluxes of components A and B, and find that the ratio is almost
constant ($\sim 0.35$) for a wide range of wavelengths. Therefore we
conclude that components A and B are two gravitationally lensed images
of a quasar at $z=1.57$. The lens hypothesis is further supported by the
fact that both components appear to have associated C \emissiontype{IV}
absorption at a blueshift of $2500{\rm km\,s^{-1}}$ from the quasar. We
note that both A and B have strong Fe \emissiontype{II}/Mg
\emissiontype{II}/Mg \emissiontype{I} absorption system at $z=1.43$, Al
\emissiontype{II} absorption system at $z=1.42$. 
This absorption is
unlikely to be associated with the lens, given the small difference
between the absorption and source redshifts. Indeed, the lensing galaxy
is typically half way out in affine or angular diameter distance, and 
probabilities of such small difference between source and lens redshifts 
are too small (e.g., \cite{ofek03}). Moreover, such absorption is
quite common since the typical linear separation between quasar
sightlines and galaxies which are responsible for such absorption is
large, $\sim 50{\rm kpc}$ \citep{bergeron91,fukugita96a}. 
It seems that strengths of absorptions are different between components
A and B, but this is not surprising given the inhomogeneity of 
the spatial distribution of absorbers (e.g., \cite{petitjean90,steidel92}). 
We note that the color of component G (see \S \ref{sec:diff}) is
consistent with that of a low-redshift ($z\lesssim 0.5$) early-type
galaxy (e.g., \cite{mcleod95}). 

\subsection{Differential Extinction\label{sec:diff}}

We check colors of quasar components to see whether the system
suffer from differential extinction. The results of photometry from the
imaging data of the SDSS, Subaru, and Keck are shown in Table
\ref{table:photo}. Relative magnitudes are estimated from Subaru, Keck,
and SDSS images. To calibrate total magnitudes, we use the measurements
of the SDSS, $i=17.26\pm0.01$, and the 2MASS, $K=15.29\pm0.18$. We find
that the colors of each component are consistent with a single value,
given the errors of $\sim 0.1 {\rm mag}$ associated with the
deconvolution of two components. Spectra of quasar components shown in
Figure \ref{fig:spec} also indicates no significant differential
extinction. Although it seems that at $\lesssim 4800${\,\AA} the ratio
increases as the wavelength decreases, but the feature is not
significant because of the large errors in those wavelengths arising
from the low response at $\sim 4000${\,\AA} or shorter. Therefore we
conclude that differential reddening is not significant.    
  
\begin{table*}
  \caption{Photometry of Subaru $i$-band, Keck
 $K$-band, and SDSS $ugriz$-band images\label{table:photo}}
  \begin{center}
    \begin{tabular}{cccccc}
     \hline\hline
      Band & Total & A & B & B$-$A & G \\
     \hline
     $u$ (SDSS) & $18.21\pm0.02$\footnotemark[$*$] & $18.54$ & $19.68$ &
     $1.14$ & $\cdots$ \\
     $g$ (SDSS) & $17.83\pm0.01$\footnotemark[$*$] & $18.12$ & $19.39$
     & $1.27$ & $\cdots$ \\ 
     $r$ (SDSS) & $17.62\pm0.01$\footnotemark[$*$] & $17.95$ & $19.09$
     & $1.14$ & $\cdots$ \\ 
     $i$ (SDSS) & $17.26\pm0.01$\footnotemark[$*$] & $17.60$ & $18.68$
     & $1.08$ & $\cdots$ \\ 
     $i$ (Subaru) & $17.26\pm0.01$\footnotemark[$*$] & $17.63$ & $18.70$
     & $1.07$ &  $20.05$ \\ 
     $z$ (SDSS) & $17.14\pm0.03$\footnotemark[$*$] & $17.49$ & $18.54$
     & $1.05$ & $\cdots$ \\ 
     $K$ (Keck)   & $15.29\pm0.18$\footnotemark[$\dagger$] & $15.78$ &
     $16.75$ & $0.97$ &  $17.81$  \\ 
    \hline
     \multicolumn{6}{@{}l@{}}{\hbox to 0pt{\parbox{85mm}{\footnotesize
       \footnotemark[$*$] Based on the SDSS.
       \par\noindent
       \footnotemark[$\dagger$] Based on the 2MASS.
     }\hss}}
\end{tabular}
  \end{center}
\end{table*}

\section{Mass Modeling\label{sec:model}}
We model the lens system with the simple Singular Isothermal Ellipsoid
(SIE) model, assuming that component G is the lens galaxy. The SIE model
has the following dimensionless surface mass density: 
\begin{equation}
  \kappa(r) = \frac{r_{\rm ein}}{2\zeta},
\end{equation}
where $\zeta=r[1+((1-q^2)/(1+q^2))\cos 2(\theta-\theta_e)]^{1/2}$,
$r_{\rm ein}$ is the Einstein ring radius, $q$ is the lens axis ratio,
and $\theta_e$ is the position angle of the ellipse. The model includes
eight parameters: the galaxy position $x_{\rm g}$ and $y_{\rm g}$, the
Einstein ring radius $r_{\rm ein}$, the ellipticity $e=1-q$, the
position angle  $\theta_e$, the source position $x_{\rm s}$ and $y_{\rm
s}$, and the flux of the quasar $f$\footnote{We obtain the same results
even if we do not adopt a parameter $f$ and instead we use only the flux
ratio, not fluxes of A and B, as a constraint.}.  Since the number of
constraints is also eight (see Table \ref{table:posflux}), the degree of
freedom is zero.  We use the public software {\it lensmodel}
\citep{keeton01b} to constrain the models.  The result is summarized in
Table \ref{table:sie}. The best-fit parameters predict the time delay
between images as $\Delta t=29.5h^{-1}{\rm day}$ ($h$ denotes the Hubble
constant in units of $100{\rm km\,s^{-1}\,Mpc^{-1}}$) if we adopt a lens
redshift $0.5$, $\Omega_M=0.3$, and $\Omega_\Lambda=0.7$. Errors on the
parameters are estimated from one-dimensional slices of the $\chi^2$
surface.  We find that we can fit the lens system with relatively small
lens ellipticity $e=0.123^{+0.041}_{-0.024}$, significantly smaller than
observed ellipticity of the light ($\sim 0.3$). The derived position
angle of the lens galaxy is consistent with that of the light
($\sim7^\circ$) within $1\sigma$.  These results are in good agreement
with previous studies (e.g., \cite{keeton98}).   

\begin{table}
  \caption{Best fit SIE model parameters for
 SDSS~J1335+0118\label{table:sie}} 
  \begin{center}
    \begin{tabular}{cc}
     \hline\hline
      Parameter & Value \\
     \hline
     $r_{\rm ein}$[arcsec] & $0.789^{+0.009}_{-0.011}$ \\
     $e$ & $0.123^{+0.041}_{-0.024}$ \\
     $\theta_e$[deg] & $-7.8^{+19.3}_{-12.0}$ \\
     \hline
     \multicolumn{2}{@{}l@{}}{\hbox to 0pt{\parbox{40mm}{\footnotesize
 We use positions and flux ratios of SDSS~J1335+0118 in the Subaru
 $i$-band image (see Table \ref{table:posflux}). Errors are 68\% 
 confidence.
     }\hss}}
\end{tabular}
  \end{center}
\end{table}

\section{Conclusion\label{sec:conc}}

We have reported the discovery of a new two-image gravitational lens
SDSS~J1335+0118. The system was identified as a new lens candidate in
the SDSS.  The image separation is $1\farcs56\pm0.002$ in the $i$-band
image taken with Subaru Suprime-cam.  The spectroscopic observation with
ESO NTT has confirmed both images have identical redshift $z=1.57$
($z=1.57\pm0.03$ and $z=1.57\pm0.05$ for components A and B,
respectively). We have also probably identified the lensing galaxy
between two quasar components in the higher-resolution imaging data
taken with Subaru Suprime-cam and Keck NIRC. The lens geometry is well
reproduced with a simple mass model and reasonable model parameters.
Assuming the redshift of the lens galaxy as $0.5$, we have predicted a
differential time delay $\Delta t\sim 30h^{-1}{\rm day}$.

\bigskip
We thank an anonymous referee for many useful comments. 
Funding for the creation and distribution of the SDSS Archive has been
provided by the Alfred P. Sloan Foundation, the Participating
Institutions, the National Aeronautics and Space Administration, the
National Science Foundation, the U.S. Department of Energy, the Japanese
Monbukagakusho, and the Max Planck Society. The SDSS Web site is
http://www.sdss.org/. 

The SDSS is managed by the Astrophysical Research Consortium (ARC) for
the Participating Institutions. The Participating Institutions are The
University of Chicago, Fermilab, the Institute for Advanced Study, the
Japan Participation Group, The Johns Hopkins University, Los Alamos
National Laboratory, the Max-Planck-Institute for Astronomy (MPIA), the
Max-Planck-Institute for Astrophysics (MPA), New Mexico State
University, University of Pittsburgh, Princeton University, the United
States Naval Observatory, and the University of Washington.

Part of the work reported here was done at the Institute of Geophysics
and Planetary Physics, under the auspices of the U.S. Department of
Energy by Lawrence Livermore National Laboratory under contract
No.~W-7405-Eng-48.   

This work is based in part on data collected at Subaru Telescope, which
is operated by the National Astronomical Observatory of Japan, and
partly based on observations collected at the European Southern
Observatory, Chile under program 70.D-0469. Some of the Data presented
herein were obtained at the W.M. Keck Observatory, which is operated as
a scientific partnership among the California Institute of Technology,
the University of California and the National Aeronautics and Space
Administration. The Observatory was made possible by the generous
financial support of the W.M. Keck Foundation. The authors wish to
recognize and acknowledge the very significant cultural role and
reverence that the summit of Mauna Kea has always had within the
indigenous Hawaiian community.  We are most fortunate to have the
opportunity to conduct observations from this mountain. We thank the
staffs of Subaru, Keck, and ESO NTT for their excellent assistance. 


\end{document}